\documentclass[a4paper,11pt]{article}
\pdfoutput=1 

\usepackage{jinstpub} 
                     
 \usepackage{graphicx}
\usepackage[caption=false]{subfig}
\usepackage{bbm}
\usepackage{comment}
\usepackage{todonotes}
\usepackage{longtable}
\usepackage[normalem]{ulem}
\usepackage{xcolor}
\usepackage{physics}
\usepackage{units}

\makeatletter
\newcommand*\dotp{\mathpalette\bigcdot@{.5}}
\newcommand*\bigcdot@[2]{\mathbin{\vcenter{\hbox{\scalebox{#2}{$\m@th#1\bullet$}}}}}

\newcommand{\pder}[2]{\frac{\partial #1}{\partial #2}}

\DeclareMathOperator{\E}{\mathbb{E}}

\DeclareMathOperator*{\argmax}{\arg\!\max}

\makeatletter
\def\diff{\@ifnextchar[{\@with}{\@without}}
\def\@with[#1]#2{\mathrm{d}^#1#2}
\def\@without#1{\mathrm{d}#1}
\makeatother
\newcommand{\define}{\equiv}
\definecolor{cardinal}{RGB}{140, 21, 21}

\title{\boldmath Parametrizing the Detector Response with Neural Networks}

\author[a,b]{S. Cheong,}
\author[a,b]{A. Cukierman,}
\author[c]{B. Nachman,}
\author[a,b,1]{M. Safdari,\note{Corresponding author.}}
\author[b]{and A. Schwartzman}

\affiliation[a]{Elementary Particle Physics Division, SLAC National Accelerator Laboratory, \\Menlo Park, CA 94025, USA}
\affiliation[b]{Physics Department, Stanford University, \\Stanford, CA, 94305, USA}
\affiliation[c]{Physics Division, Lawrence Berkeley National Laboratory, \\Berkeley, CA 94720, USA}

\emailAdd{murtazas@stanford.edu}

\abstract{
In high energy physics, characterizing the response of a detector to radiation is one of the most important and fundamental experimental tasks.  In many cases, this task is accomplished by parameterizing summary statistics of the full detector response probability density.  The parameterized detector response can then be used for calibration as well as for directly improving physics analysis sensitivity.   This paper discusses how to parameterize summary statistics of the detector response using neural networks.  In particular, neural networks are powerful tools for incorporating multidimensional data and the loss function used during training determines which summary statistic is learned.  One common summary statistic that has not been combined with deep learning (as far as the authors are aware) is the mode.  A neural network-based approach to mode learning is proposed and empirically demonstrated in the context of high energy jet calibrations.  Altogether, the neural network-based toolkit for detector response parameterization can enhance the utility of data collected at high energy physics experiments and beyond.
}

\keywords{Analysis and statistical methods; Data processing methods; Large detector-systems performance; Pattern recognition, cluster finding, calibration and fitting methods.}

\arxivnumber{1910.03773} 

\begin{document}
\maketitle
\flushbottom

\section{\label{sec:intro}Introduction}

Any physics experiment is inevitably influenced by one's ability to detect and measure physical observables of interest. In particular, an accurate understanding of the characteristics of detectors is important for designing an experiment, projecting the reach of an experiment, calibrating measurements, and ultimately analyzing the data collected. Therefore, one of the most fundamental and important experimental tasks is to understand the behavior of the detection apparatus.   Beyond the physical energy deposited by radiation, measurements from real physics experiments are influenced by the electronic read-out processes as well as a variety of other hardware-~, firmware-, and software-based algorithms.  All of these effects are collectively called \textit{the effect of a detector}.

Mathematically, any experimental measurement can be considered as a function that intakes some random variable $X$ and outputs another random variable $Y$.  The input is the `true' quantity and the output is the `measured' quantity.   Due to non-negligible noise\footnote{Some of this noise is from marginalizing out unmeasured quantities while other noise is quantum-mechanical in origin.}, $Y|X$ is also a random variable and its distribution can be described by the probability distribution $p(y|x)$, where $x$ and $y$ are realizations of $X$ and $Y$.  Because the detector response is captured completely by the distribution $p(y|x)$, this distribution is referred to as \textit{the response}.

For example, $X$ could be the true energy of a particle traversing the detector, while $Y$ is the reconstructed energy measured by the detector. $X$ and $Y$ need not have the same physical dimensions; for instance, a tracking detector immersed in a magnetic field aims to determine the momentum $X$ of a charged particle by making spatial location measurements $Y$ and calculating the sagitta of the particle trajectory. It is also worth noting that both $X$ and $Y$ are often multi-dimensional; that is, a detector may measure multiple observables simultaneously and its response may depend on multiple variables including ones other than observables of primary interest.

For calibration and other experimental tasks, it is often necessary to parametrize detector effects with a function $\nu(x)$ that categorizes the response of the detector to an input $x$.  The function $\nu$ encodes some summary statistic(s) of the probability distribution $p(y|x)$ and often represents the central tendency of the detector response such as the mean or the spread of the response such as the standard deviation.  The parameterization $\nu(x)$ is typically obtained using detailed Monte Carlo (MC) simulations of an experiment or calibration data where the values of the input variable $x$ are measured precisely by independent techniques.

A common method to obtain $\nu(x)$ is to bin the (simulated) data in $X$ and then calculate $\nu$ based on a histogram of $Y|X$.  This method provides a ``look-up" table for values of $\nu(x)$ at the center of each bin of $X$, which can then be interpolated.   While this method works well when $X$ is low dimensional, it becomes infeasible when more observables are added to improve the accuracy of the detector description.  In order to maintain the same statistical precision in a given $X$ bin, the bin sizes would need to grow exponentially large for a linear increase in the dimensionality of $X$.  Furthermore, interpolation between bins is more difficult in higher dimensions.  The task is additionally challenging if $Y$ is multidimensional.

An alternative to the bin method is to perform an unbinned fit using a functional form for $p(y|x)$ and $\nu(x)$.  Such functional forms are nearly always an approximation and often do not capture all of the salient features of the data.  For example, it is common to assume that $p(y|x)$ has a Gaussian core even though the tails are known to not be simply Gaussian. If the assumptions about $p(y|x)$ or $\nu(x)$ are inaccurate, any statistical fit will result in a biased parametrization of the detector, which will influence all consequent experimental studies.  Hybrid methods using a functional form for $p(y|x)$ and a non-parameteric form for $\nu(x)$ are also possible (see e.g.~\cite{Khachatryan:2015iwa,Aaboud:2018ugz}).

This paper describes methods for parameterizing the detector response using neural networks (NN's).  By their design, NN's can readily incorporate unbinned high-dimensional inputs and outputs. The form of $\nu(x)$ will dictate the loss function used in the NN training.  This connection will be explained in detail for the mean, median, standard deviation, and quantiles of $p(y|x)$.  A new method (as far as the authors are aware) is proposed for a NN parametrization when $\nu(x)$ is the mode of $p(y|x)$.  

This paper is structured as follows. A mathematical overview of NN training is provided in Sec.~\ref{sec:math}. Then, Sec.~\ref{sec:response} discusses how to learn the various properties and obtain a parametrization $\nu(x)$ of the detector's response using NN's. In Sec.~\ref{sec:examples}, the proposed method is numerically demonstrated using realistic examples of a physical detector. Sec.~\ref{sec:discussion} discusses the potential uses of the proposed method across experimental physics and a further application to neural-network-based calibration of detectors. The paper ends with conclusions and future outlook in Sec.~\ref{sec:concl}.


\section{\label{sec:math}Mathematical Overview of Training Neural Networks}

Neural network are trained by minimizing a loss function $\ell$: 

\begin{align}
\label{eq:learning}
\nu&=\text{argmin}_{\nu'}\mathbb{E}[\ell(Y,\nu'(X))]\\\noindent 
&=\text{argmin}_{\nu'}\iint \ell(y, \nu'(x))\, p(x, y) \dd{x} \dd{y},
\end{align}

\noindent where $\mathbb{E}[\cdot]$ represents the expected value or average.  The form of $\ell$ controls what statistic of $p(y|x)$ will be parameterized by $\nu$.  One of the most common loss functions is the squared error $\ell(\nu(x),y)=\norm{y-\nu(x)}_2^2$.  Loss functions must be bounded below and typically have a unique local minimum.  As it is the most common case and significantly simplifies the notation, the following discussion will focus on the case where $Y$ is one-dimensional.

If the training dataset consists of a large enough unbiased sample from the distribution $p(x, y)$, then a sufficiently flexible NN and training procedure will arrive at Eq.~\ref{eq:learning}. Consider a small neighborhood $V(x_0)$ around some $x_0$ within the support of the training set. Assuming that the distribution $p(x, y)$ and the value of the NN function $\nu_0 \define \nu(x_0)$ are roughly constant (vary slowly) within $V(x_0)$, a successfully trained NN must satisfy:

\begin{equation}
\label{eq:nn_train_condition}
\begin{aligned}
	0	&= \pder{}{\nu_0}\mathbb{E}[\ell(Y,\nu(X))]	\\
		&= \pder{}{\nu_0}
			\left[ \int_{V(x_0)} p(x) \int_{-\infty}^{\infty} \ell(y, \nu(x)) p(y|x) \dd{y} \dd{x} \right]		\\
		&\approx
			\left[ \int_{V(x_0)} p(x) \dd{x} \right] \times
			\left[ \int_{-\infty}^{\infty} \pder{\ell(y, \nu_0)}{\nu_0} p(y | x_0) \dd{y} \right]
			\\
		&\propto
			\E \left[ \left. \pder{\ell(y, \nu_0)}{\nu_0} \right| x_0 \right].
\end{aligned}
\end{equation}

\noindent As $\ell$ is bounded from below, Eq.~\ref{eq:nn_train_condition} must be true everywhere across the support of $p(x)$ for a properly trained NN.  Therefore, $\E\left[ \left. \partial \ell(y, \nu(x)) / \partial \nu(x) \right| x \right] = 0$ which suggests that one can therefore engineer $\ell$ to target a statistic of interest for the response parameterization.  

\section{\label{sec:response}Learning Different Properties of the Response Distribution}

The sections below discuss loss functions that can be used to learn some common statistics for response parametrization using NN's.  The specific choice depends on the particular experimental application.  Hybrid methods are also available, such as the Huber loss~\cite{huber1964} which interpolates between the mean squared error and the mean absolute error.  The asymptotic behavior of such approaches are more complicated than the ones presented in this section, but they may be useful in practice for certain situations.


\subsection{\label{subsec:mean}Mean}

The mean of the response distribution $p(y | x)$ can be learned using the mean square-error (MSE) loss:  $\ell_\text{SE}(y, \nu(x)) \define (y-\nu(x))^2$.  According to Eq.~\ref{eq:nn_train_condition}, a NN trained successfully with $\ell_\text{SE}$ will satisfy:
\begin{equation*}
	\E\left[ \left. \pder{\ell_\text{SE}(y, \nu(x))}{\nu(x)} \right| x \right]
		= 2\left( \nu(x) - \E[y | x] \right) = 0.
\end{equation*}

\subsection{\label{subsec:median}Median}

The median may be a useful alternative to the mode for characterizing the central tendency of $Y|X$ because it is less sensitive to outliers.  The median of the response can be learned using the mean absolute-error (MAE) loss:  $\ell_\text{AE}(y, \nu(x)) \define |y-\nu(x)|$.  A NN trained with MAE will output $\nu(x)$ such that:
\begin{equation}
	\E\left[ \left. \pder{\ell_\text{AE}(y, \nu(x))}{\nu(x)} \right| x \right]
		= \E \big[ \mathrm{sgn}(y - \nu(x)) \big| x \big] = 0.
	\label{eq:mae_train_condition}
\end{equation}
where $\mathrm{sgn}(z)$ is the signum function defined by:
\begin{equation*}
	\mathrm{sgn}(z) \define
	\begin{cases}
		1		&	\text{if $z > 0$}	\\
		0		&	\text{if $z = 0$}.	\\
		-1		&	\text{if $z < 0$}
	\end{cases}
\end{equation*}
Then, Eq.~\ref{eq:mae_train_condition} can be more explicitly written as:
\begin{equation*}
	-\int_{-\infty}^{\nu(x)} p(y | x) \dd{y} + \int_{\nu(x)}^{\infty} p(y|x) \dd{y} = 0.
\end{equation*}
which precisely requires that $\nu(x)$ be the median of the distribution $p(y|x)$.

\subsection{\label{subsec:mode}Mode}

Like the median, the mode is insensitive to outliers.  The mode has the additional property that it is insensitive to truncating the distribution on the left or right side, as is often necessary for realistic detector readout systems.  However, unlike the median, the mode may not be unique.  The mode of the response can be learned with  $\ell(y, \nu(x)) = -\delta(y-f(x))$, where $\delta$ is the Dirac delta function.   The solution to Eq.~\ref{eq:nn_train_condition} is shown using the properties of the Delta-function:

\begin{equation*}
	-\int_{-\infty}^{\infty} \pder{\delta(y - \nu(x))}{\nu(x)} p(y | x) \dd{y}
		= -\left. \pder{p(y|x)}{y} \right|_{y = \nu(x)} = 0.
\end{equation*}

The Dirac delta-function is an impractical loss function and so one must use a surrogate loss with similar properties.   Various surrogates have been studied in the context of linear regression~\cite{moderegression1,moderegression2}, but as far as the authors are aware, this has never been combined with neural networks.

One possible surrogate to the Dirac delta-function is a negative Gaussian kernel:
\begin{equation*}
	\ell_\text{GK}(y, \nu(x); h) \define
		-\frac{1}{\sqrt{2\pi}h} \exp(-\frac{(y - \nu(x))^2}{2 h^2})
\end{equation*}
where $h$ is a hyper-parameter called the ``bandwidth" of the kernel.

Using a (negative) kernel function as a loss function to learn the mode is analogous to the core idea of kernel density estimation (KDE). KDE estimates a probability distribution by finite sum of kernels centered at a set of data points. Analogously, the mean training loss in a small neighborhood of given $x$ using $\ell_\text{GK}$ will reproduce a smeared version of $-p(y|x)$ up to a normalization factor and a constant-offset. Therefore, the mean training loss is minimized when $\nu(x) = \argmax_y p(y|x)$ across the support of the training dataset.

\begin{figure}[t]
	\centering
	\subfloat[Gamma Distribution with $k=2$ and $\theta=1$ \label{fig:mode_loss_landscape_gamma}]{
		\includegraphics[width=0.45\textwidth]{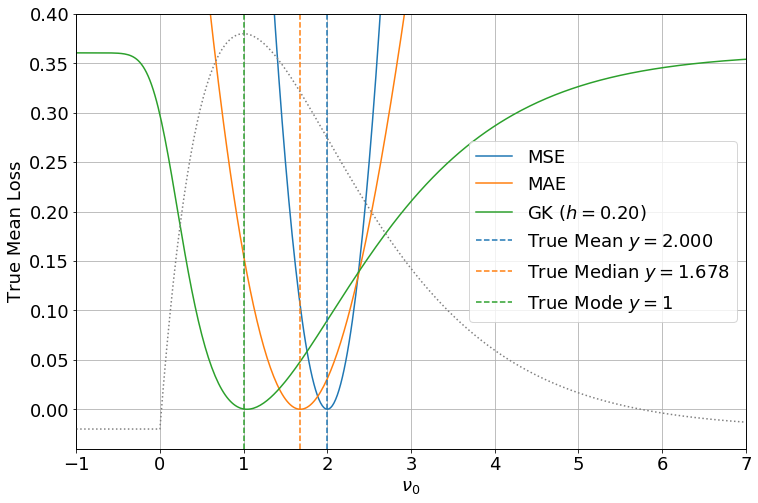}
	}
	\subfloat[Crystall Ball Distribution with $\beta=1$ and $m=2$ \label{fig:mode_loss_landscape_CB}]{
		\includegraphics[width=0.45\textwidth]{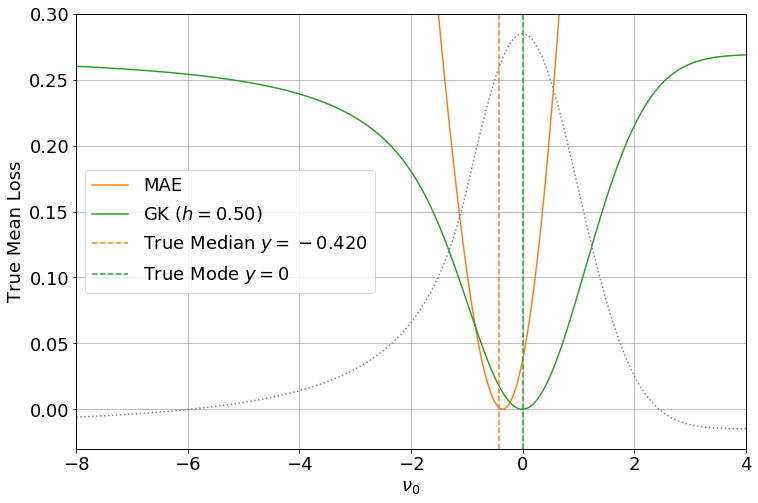}
	}\\
	\caption{The true mean losses as a function of $\nu_0$ for two different underlying distributions, each with a unique mode.
		Shown in gray dots are the underlying probability distributions, with arbitrary vertical scale. Constant off-sets have been subtracted from the loss functions such that their minima are all zero (purely for plotting purposes).}
	\label{fig:mode_loss_landscape}
\end{figure}

The potential of $\ell_\text{GK}$ is illustrated in Fig.~\ref{fig:mode_loss_landscape}, which shows the loss (analytically as in Eq.~\ref{eq:learning}) landscape for a fixed $x=x_0$.

Plotted in Fig.~\ref{fig:mode_loss_landscape} are the true mean losses for a given distribution of $y$:
\begin{equation}
	\mathbb{E}[\ell(Y, \nu_0)|X=x_0]=\int_{-\infty}^{\infty} \ell(y, \nu_0) p(y|x_0) \dd{y},
	\label{eq:mean_loss_at_point}
\end{equation}
where $\nu_0=\nu(x_0)$.  For both the Gamma and Crystal Ball distributions, the mean GK loss mimics a smoothened version of the underlying probability density distribution.   The GK loss does not reproduce the strict cut-off of the Gamma distribution at $y=0$, but its minimum is still unbiased with respect to the true mode.  As long as $p(y|x)$ is smooth around the mode, the function minimized with the GK loss predicts the mode.  As advertised in Sec.~\ref{subsec:mean} and~\ref{subsec:median}, Figure~\ref{fig:mode_loss_landscape} shows that the minima of MSE and MAE losses are at the mean and the median of $p(y|x)$, respectively.  For such asymmetric response distributions, the mode is a good parameter with which to characterize the typical response of a detector, especially since the mean does not exist for the Crystal Ball distribution.

The negative Gaussian kernel may also not be a practical loss because its gradients vanish away from the mode, reducing the effectiveness of gradient-based optimization methods.  One solution to this challenge is to introduce ``leakiness" to the negative Gaussian kernel by defining the leaky Gaussian kernel (LGK) loss:
\begin{equation}
	\ell_\text{LGK}(y, \nu(x); h, \alpha)
		= \ell_\text{GK}(y, \nu(x); h) + \alpha \abs{y - \nu(x)},
	\label{eq:lgk_loss}
\end{equation}
where $\alpha$ is a hyper-parameter called the leakiness. Note that the loss is still minimized (zero) when $\nu(x) = y$.  The leakiness should be $\alpha\ll 1$ so that the value of $\nu(x)$ that solves Eq.~\ref{eq:nn_train_condition} is still close to the mode of $p(y|x)$. This term ensures a lower bound on the magnitude of the loss gradient, even when the NN output is far away from the optimal target value.

\begin{figure}[h]
	\centering
	\includegraphics[width=0.45\textwidth]{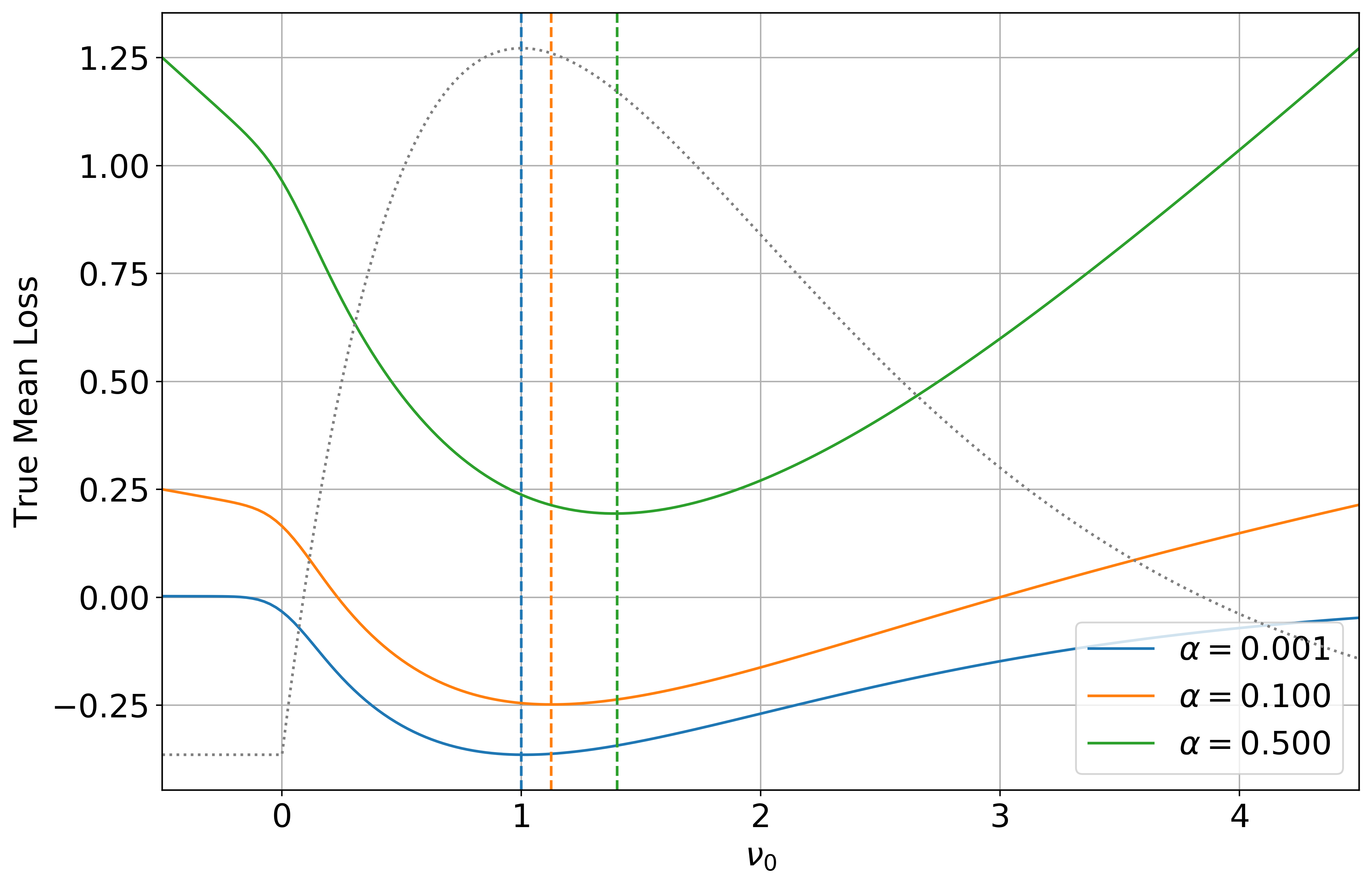}
	\caption{The true mean LGK losses with different $\alpha$'s for the gamma distribution. The minimum of each mean loss function is shown with a dotted vertical line. The true mode of the distribution is at $y=1$, and its median at $y \approx 1.678$.}
	\label{fig:mode_loss_alpha}
\end{figure}

Fig.~\ref{fig:mode_loss_alpha} shows the true mean losses (Eq.~\ref{eq:mean_loss_at_point}) with various values of $\alpha$ for the gamma distribution with $k=2$ and $\theta=1$. Note that the mean LGK loss is minimized at values very close to the true mode (which is at $y=1$) for small values of $\alpha$. However, when $\alpha$ is large, the LGK loss approaches the absolute error $\ell_\text{AE}(y_1, y_2) \define \abs{y_1 -y_2}$, and the minimum therefore gets biased towards the median at $y \approx 1.678$. Hence, as far as $\alpha$ is small, the leakiness term can improve the training of NN's trying to learn the mode via the LGK loss, while introducing only an $\alpha$-suppressed bias.

%

\subsection{\label{subsec:stdev}Standard Deviation}
The three parameters mentioned so far are useful for characterizing the central tendency (i.e. some representative value) of the detector's response. However, also important for characterizing the detector are measures of the dispersion of the response distribution. Such quantities allow for parametrization of the precision of the detector's measurements.

For example, one may choose to characterize the dispersion of a detector's response using the standard deviation $\sigma(y | x)$ of its response. This can be parametrized using two NN's each trained using $\ell_\text{MSE}$.  In particular, a modified dataset can be combined with $\ell_\text{MSE}$ to learn any moment of the distribution of $p(y|x)$.  This is accomplished by training with examples drawn from $X$ and $Y^n$ instead of the usual $X$ and $Y$.  One network trained with $n=1$ (same as Sec.~\ref{subsec:mean}) and another network trained with $n=2$ can be combined to parametrize the standard deviation of the response:
\begin{equation*}
	\sigma(y | x) \define \sqrt{\E[y^2 | x] - \E[y|x]^2}.
\end{equation*}

While a network trained to learn the second moment of $y$ works well (similarly to learning the mean), the standard deviation estimated in this way is unstable because it requires taking the sum in quadrature of two estimates.  Numerical instabilities in the required fine cancellation of large numbers results in a poor estimate of $\sigma(y|x)$.

\subsection{\label{subsec:quantile}Quantile}
Another useful measure for the dispersion of the detector response is the quantile. As with the median, the advantage of statistics like the interquartile range over the standard deviation is the reduces sensitivity to outliers.  A particular quantile of the response distribution can also be used to parametrize the upper/lower limits on the detector's output.


One can learn arbitrary quantiles of $p(y|x)$ by extending the absolute error $\ell_{AE}$ to the generalized absolute-error (GAE)~\cite{pinball3,pinball2}:
\begin{equation}
	\ell_\text{GAE}(y, \nu(x)) =
	\begin{cases}
		q\abs{y - \nu(x)}		&	y \geq \nu(x)		\\
		(1-q)\abs{y - \nu(x)}	&	y < \nu(x)	
	\end{cases}
\end{equation}
where $q \in [0, 1]$ represents the quantile that one wants to learn.  This can also be extended to be smooth~\cite{pinball4,pinball5}, simultaneously learn multiple quantiles~\cite{pinball}, as well as other variations and extensions.

The mean training loss using GAE is minimized when:
\begin{equation*}
	\frac{\displaystyle \int_{\nu(x)}^{\infty} p(y|x) \dd{y}}
		{\displaystyle \int_{-\infty}^{\nu(x)} p(y|x) \dd{y}}
	= \frac{1-q}{q}
\end{equation*}
which precisely requires that $\nu(x)$ is the $100q \%$ quantile of the distribution $p(y|x)$. 

\section{\label{sec:examples}Numerical Examples}

This section uses two examples from collider high-energy physics in order to illustrate the concepts presented in earlier sections.

Monte Carlo samples $\{(x_i, y_i)\}$ are generated following distributions $p(y|x)$ that are motivated by realistic detectors.  In particular, the first example approximates the calorimeter response of a hadronic jet and the second example is motivated by the momentum reconstruction of muons.  The jet energy is used to demonstrate the power of mode learning and the muon momentum reconstruction illustrates the potential of quantile learning.  In both cases, the prior distribution $p(x)$ is chosen to be uniform and the goal is to compare the mode or quantiles to the analytic result based on $p(y|x)$.



\subsection{\label{subsec:jet_energy}Calorimeter and Jet Energy Response}

The first example is a calorimeter used to measure hadronic jets in collider experiments. Jets are collimated sprays of particles resulting from quarks and gluons produced at high energy.   Details of the ATLAS and CMS jet calibrations can be found in Ref.~\cite{Aad:2011he,Aad:2014bia,Aaboud:2017jcu,Khachatryan:2016kdb,Chatrchyan:1369486}.


Because the calorimeter measures the energy of a large number of particles produced in a hadronic particle shower, the jet energy response is approximately Gaussian. However, jet energy measurements also typically demonstrate non-Gaussian tails due to various detector effects such as inactive material, punch-through beyond the calorimeter, etc.  In addition, jet reconstruction algorithms often place a minimum transverse energy threshold to reduce the high rate of noise jets resulting from calorimeter noise, random combinations of low-energy particles, or jets from additional nearly simultaneous collisions (pileup).  To illustrate the impact of these effects, the example considered in this section uses a truncated Crystal Ball distribution:

\begin{equation}
	p(y | x) =
	\begin{cases}
		\displaystyle 0					&	\text{if } y < E_\text{min}	\\
		\displaystyle A_1 \left( \frac{m}{\beta} - \beta - \frac{y - \mu(x)}{\sigma(x)} \right)^{-m}	\\
		&	\hspace{-1.2cm}	\text{if } E_\text{min} < y < E_\text{tail}		\\
		\displaystyle A_2 \exp(-\frac{(y - \mu(x))^2}{2 \sigma(x)^2})	&\text{if } y > E_\text{tail}
	\end{cases}
	\label{eq:jet_response_dist}
\end{equation}
where $x$ is the true jet energy, $y$ is the reconstructed jet energy, $E_\text{min}$ is the minimum jet energy imposed by the reconstruction algorithm, and $A_{1,2}$ are normalization factors. $E_\text{min}$ is set to 10 in this study. $\beta$, $m$, and $E_\text{tail} \define \mu(x) - \beta \sigma(x)$ are parameters of the Crystal Ball distribution. $\beta$ and $m$ are set to $1.5$ and $2$, respectively. These values are chosen such that the threshold effects and the non-Gaussian tails have a non-trivial impact on the mean energy response.  The units are chosen to be typical for a calorimeter measurement in GeV.

The parameters $\mu(x)$ and $\sigma(x)$ characterize the central tendency and spread of the jet response.  The reconstructed jet energy is typically smaller than the true jet energy because hadronic calorimeters are composed of layers of active and inactive (absorber) material and the latter do not collect any signal.  For illustration, the functional form of $\mu$ is taken from Ref.~\cite{Cukierman:2016dkb} and motivated by the ATLAS and CMS calorimeter jet energy response:

\begin{equation}
	\mu(x) = x + 5\ln(e^{x_0 / 10} + e^{x / 10}) - 5\ln(1 + e^{(x + x_0) / 10}),
	\label{eq:jet_response_mu}
\end{equation}
where $x_0$ is a turn-on parameter for the response function, set to $25$ in this toy example.  The jet energy resolution function $\sigma(x)$ uses the a typical parameterization taking into account a noise term ($N$), stochastic term ($S$), and constant term ($C$):
\begin{gather}
	\sigma(x) = 
		N \oplus S\sqrt{x} \oplus Cx,
	\label{eq:jet_response_sigma}
\end{gather}
where $N$ represents energy-independent effects such as detector noise and pileup interactions, $S \sqrt{x}$ represents the Poisson-distributed, statistical fluctuations, and $Cx$ represents fluctuations proportional to the true jet energy $x$, such as energy lost in the passive absorber volume.  For the example presented here, $N=4$, $S=0.6$, and $C=0.05$.  These values roughly reproduce the transverse energy resolutions reported by ATLAS and CMS, with a relative resolution of approximately 25\% (10\%) at transverse energy of $\unit[20]{GeV}$ ($\unit[100]{GeV}$).

\begin{figure}[t]
	\centering
	\includegraphics[width=0.45\textwidth]{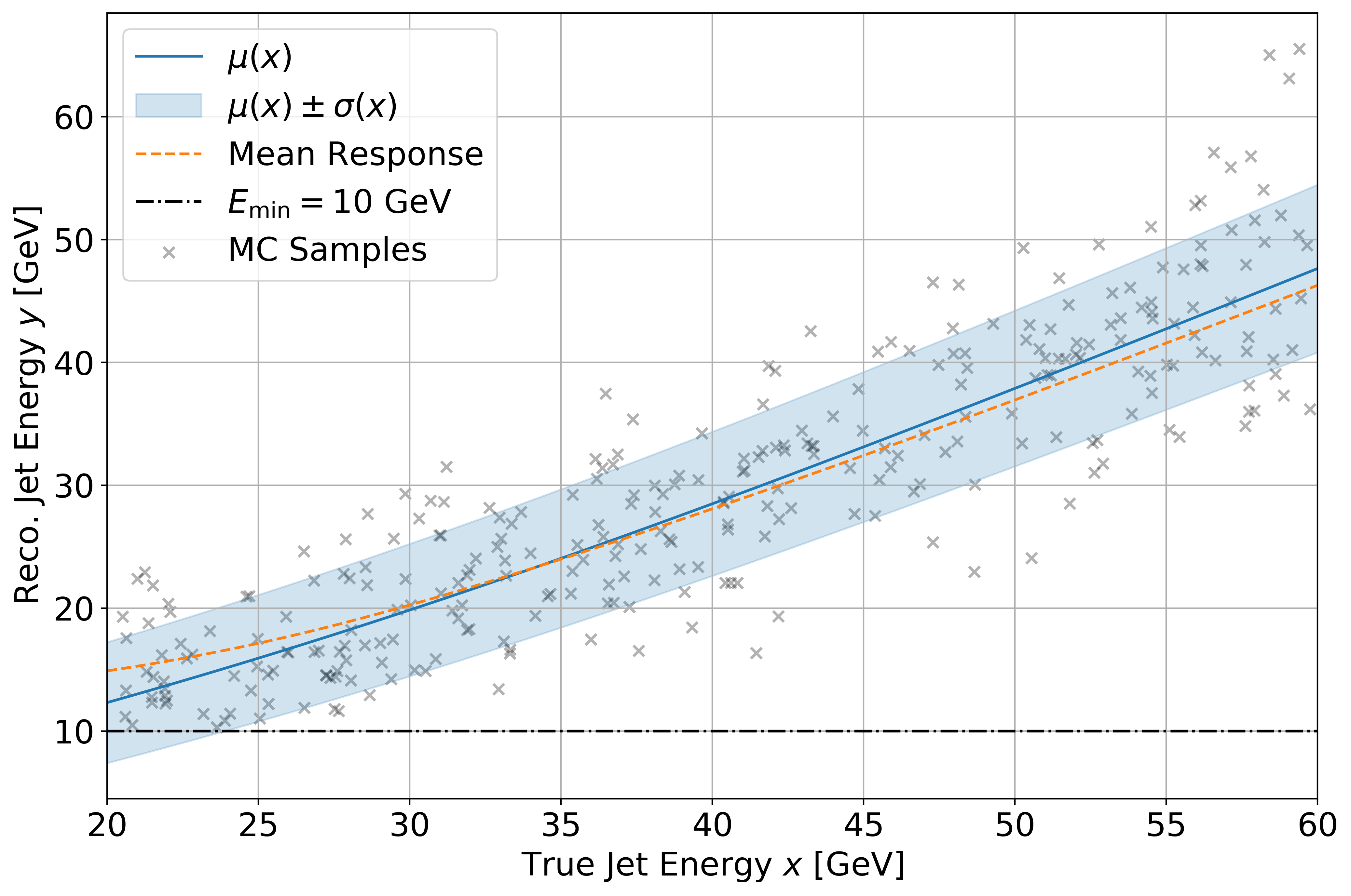}
	\caption{Jet energy response similar to those at ATLAS and CMS. 300 random MC samples are plotted together.}
	\label{fig:jet_response_distribution}
\end{figure}

For training, 500,000 examples are generated using a uniform distribution in $x$ and then the response distribution defined by Eqs.~\ref{eq:jet_response_dist}, \ref{eq:jet_response_mu}, and~\ref{eq:jet_response_sigma}.  These data are shown in Fig.~\ref{fig:jet_response_distribution} alongside the analytic values of $\mu$ and $\sigma$. Note that some low energy jets are lost because of the minimum energy cut at $E_\text{min} = \unit[10]{GeV}$, particularly below $x \approx \unit[30]{GeV}$.   As a result, the sample mean response is above the true value of $\mu(x)$.  In the high-energy regime, threshold effect diminishes, and the non-Gaussian tail causes the mean response to be lower than $\mu(x)$. Therefore, if one were to characterize the jet energy response of the calorimeter, the mean response may not be an appropriate choice of parametrization.

To show the benefit of mode learning, two NN's are trained using the same dataset, one with the MSE loss (for the mean, Sec.~\ref{subsec:mean}) and one with the LGK loss (for the mode, Sec.~\ref{subsec:mode}).  The inputs to the networks are the true jet energy $x$ and an additional feature $x^3$, added to ease the learning of a non-linear function. The two neural networks have identical architectures: an input layer with two nodes, a single hidden layer with $50$ nodes with parametric rectified linear unit (PReLU)~\cite{2015arXiv150201852H} as activation functions, and an one-dimensional output layer. Both networks are implemented in Keras~\cite{keras} using the Tensorflow backend~\cite{tensorflow} and trained using the \textsc{Adam} optimization method~\cite{DBLP:journals/corr/KingmaB14}, until their mean training losses converge. This process is repeated 25 times to obtain a set of 25 NN's parametrizing the mean and another set parametrizing the mode.

\begin{figure}[t]
	\centering
	\subfloat[\label{fig:jet_response_all_NNs}]{
		\includegraphics[width=0.45\textwidth]{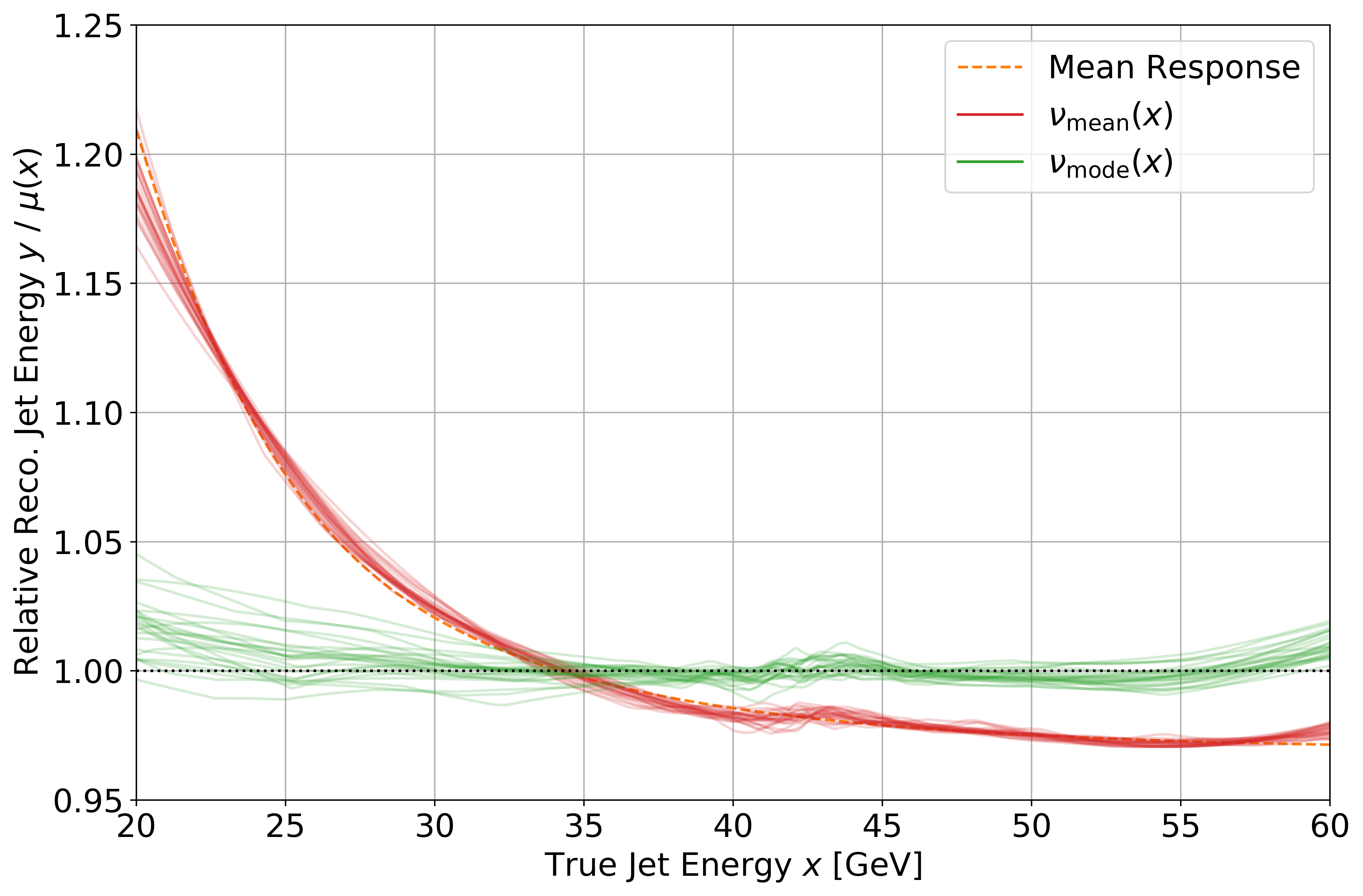}
	}
	\subfloat[\label{fig:jet_response_param_results}]{
		\includegraphics[width=0.45\textwidth]{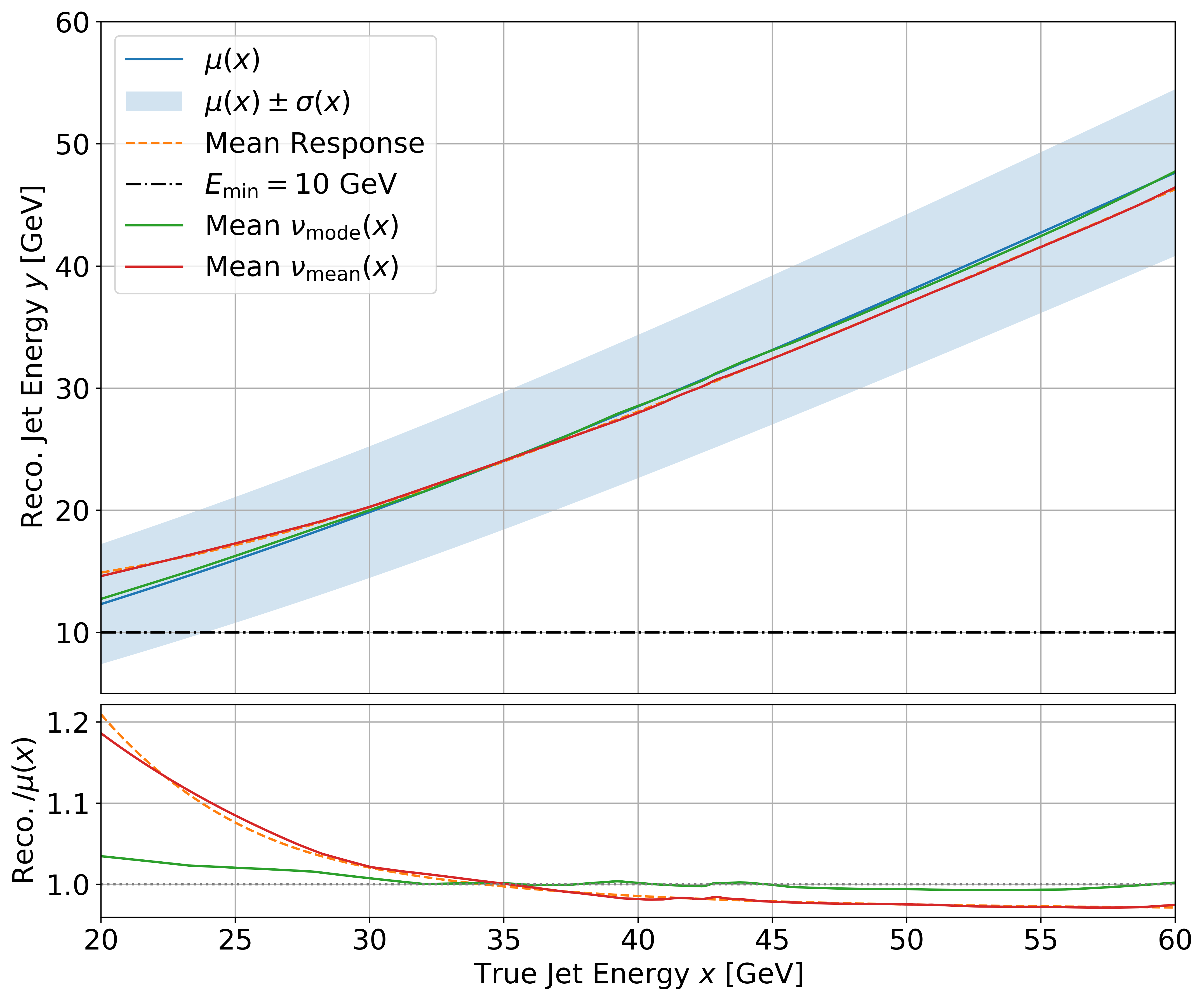}
	}
	\caption{(a) All NN's trained to parametrize the jet energy response: 25 $\nu_\text{mean}$'s and $\nu_\text{mode}$'s each. $\nu_\text{mean}$'s are trained with the MSE loss, and $\nu_\text{mode}$'s are trained with the mean LGK loss with $h=0.1$ and $\alpha = 10^{-4}$. (b) True parameters and NN-parametrizations of the jet response. The red and green lines show the mean of 25 $\nu_\text{mean}$'s and $\nu_\text{mode}$'s respectively.}
\end{figure}

Fig.~\ref{fig:jet_response_all_NNs} shows the final results obtained from all the NN's. The $\nu_\text{mode}$'s trained with the LGK loss generally follow the mode of the response $\mu(x)$, while the $\nu_\text{mean}$'s trained with the MSE loss follows the mean response (distorted from $\mu(x)$). To reduce variance from initialization and training processes, the average of 25 NN's of each type are combined. These results are shown in Fig.~\ref{fig:jet_response_param_results}. The mode-learning NN's are clearly more appropriate, if one were to parametrize the jet energy response robust to the minimum-energy cutoff and the non-Gaussian tails.

\subsection{\label{subsec:silicon_sensor}Muon Momentum Resolution}

This second example is used to illustrate the power of quantile learning to characterize the spread of the response distribution.   For this purpose, muon momentum reconstruction provides the motivation.  Characterizing the muon momentum resolution is important for resonance searches~\cite{Aad:2019fac,Sirunyan:2018exx} and has both Gaussian and non-Gaussian components.  Further details of the muon momentum reconstruction can be found in Ref.~\cite{Aad:2016jkr,Aad:2014rra,Chatrchyan:2012xi,Sirunyan:2018fpa}.  

As in the previous example, a Crystal Ball distribution is used to model the resolution for the reconstructed muon momentum.  In particular, $y|x=1+\sigma(x)z$ where $z$ follows a standard Crystal Ball distribution (Eq.~\ref{eq:jet_response_dist} with $\mu=0,\sigma=1,E_\text{min}=-\infty$) with $m=2$.  The resolution function is chosen to approximately match the ATLAS and CMS momentum resolution:

\begin{align}
\sigma(x)=0.03\left(\frac{1}{4}e^{20\log_{10}(\log_{10}(x))-4}+1\right),
\end{align}

\noindent where $x$ is the momentum, given in units of GeV.  The distribution of $\log_{10}(x)$ is sampled uniformly for training between $\log_{10}(7)$ and 2.  This distribution has the property that it is well-described by a Gaussian when $|x-1|$ is smaller than the Crystal Ball $\beta$ parameter and otherwise poorly described by a Gaussian.  Furthermore, the heavy tail makes the standard deviation diverge.  The goal is to show that learning the $68\%$ inter-quantile range is a robust estimate of the core Gaussian standard deviation and is insensitive to the heavy tails governed by $\beta$. 

\begin{figure}[t]
	\centering
	\subfloat[\label{fig:muon_response_all_NNs}]{
		\includegraphics[width=0.45\textwidth]{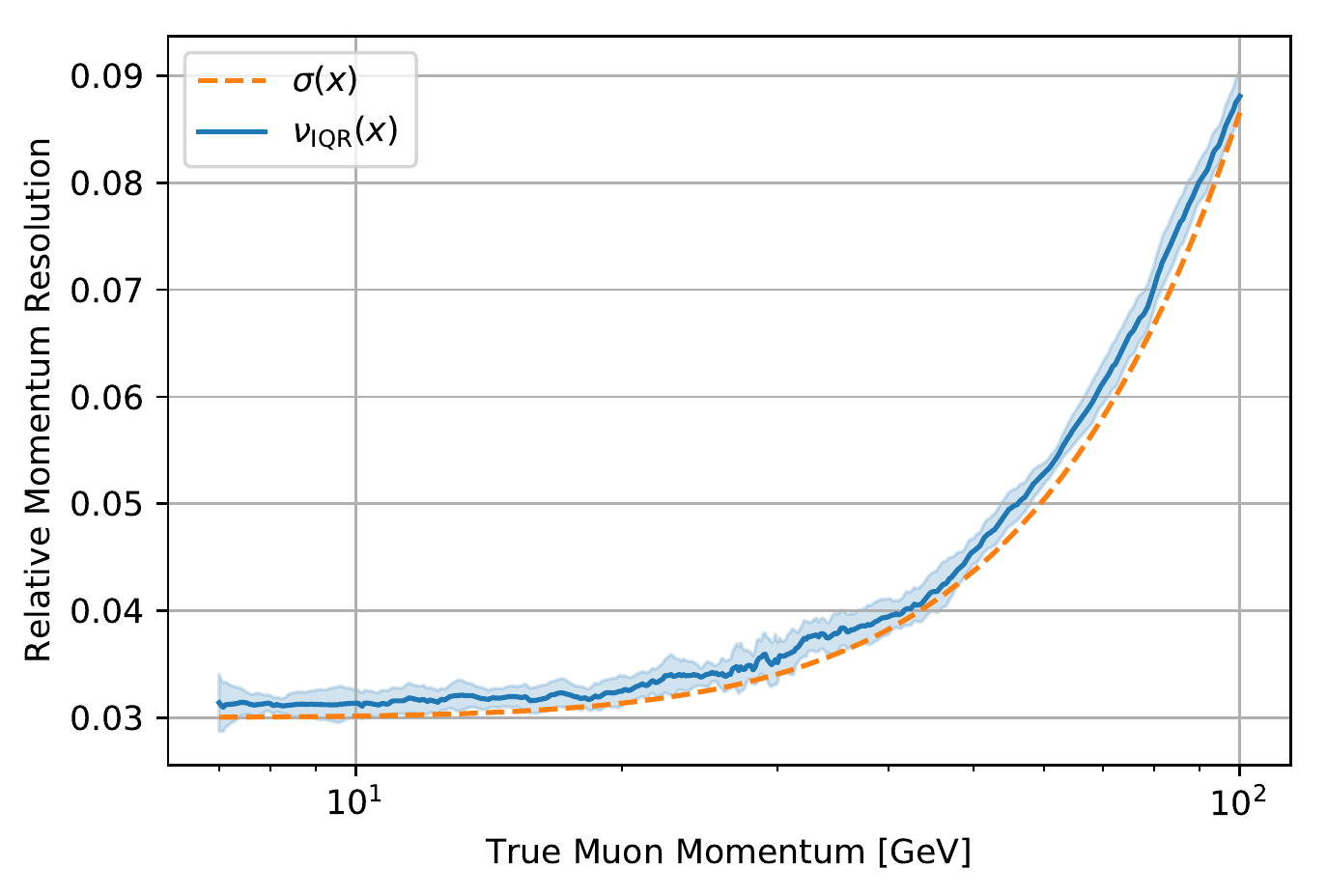}
	}
	\subfloat[\label{fig:muon_response_param_results}]{
		\includegraphics[width=0.45\textwidth]{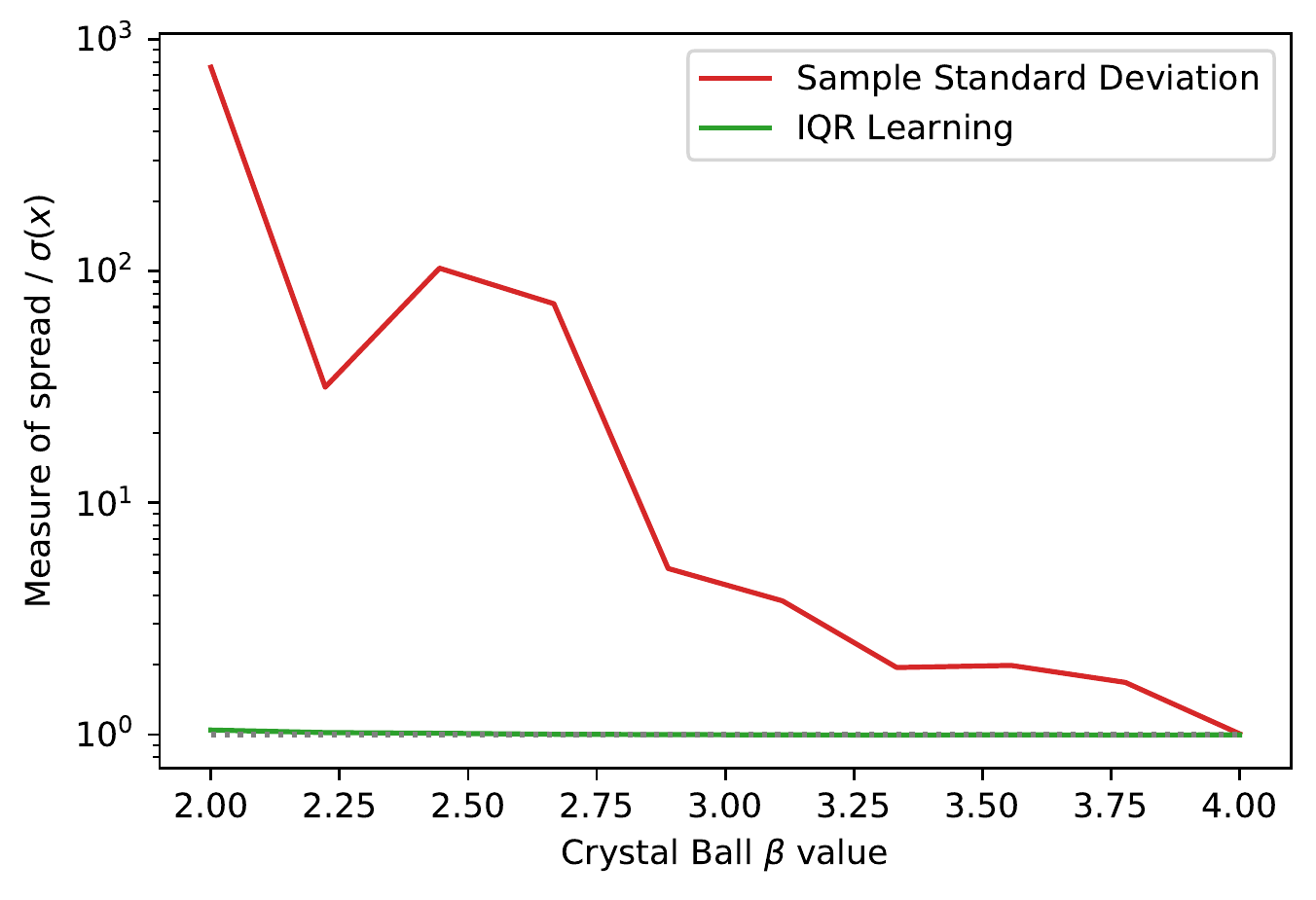}
	}
	\caption{(a) The median (central value) and 68\% inter-quantile range (error band) using the 25 NN's trained to parametrize the 68\% inter-quantile range of the muon momentum response distribution with $\beta=2$.  The NN's are optimized using the GAE loss.  (b) The measured spread divided by $\sigma(x)$ as a function of $\beta$.  For the sample standard deviation, this is the sample standard deviation of $(y-1)/\sigma(x)$ over all $(x,y)$ values.  The corresponding quantity for the NN is the average value of $\frac{1}{2}(\nu_\text{GAE,$q=0.16$}(x)-\nu_\text{GAE,$q=0.84$}(x))/\sigma(x)$.}
\end{figure}

To learn the 68\% inter-quantile range, two neural networks are trained to learn the 84\% and 16\% quantiles using the GAE loss (Sec.~\ref{subsec:quantile}).  Half of the difference between these NN outputs should approximate the Gaussian core standard deviation.  Figure~\ref{fig:muon_response_all_NNs} shows the learned value of the 68\% inter-quantile range compared with $\sigma(x)$ as a function of $x$ for $\beta=2$.  Even though there are heavy tails, the quantile regression is able to accurately model the core Gaussian standard deviation.  Figure~\ref{fig:muon_response_param_results} demonstrates that the GAE loss produces an estimate that is robust to variations in $\beta$.  Smaller values of $\beta$ correspond to heavier tails.  For comparison, the sample standard deviation is shown alongside the estimate of the NN to show the impact of the heavy tails on a naive estimate of the standard deviation.  As $\beta\rightarrow\infty$, the heavy tails are suppressed and the sample standard deviation approaches the core Gaussian standard deviation.

\section{\label{sec:discussion}Discussion}

NN's can provide data-driven parameterizations of detector responses, free of \textit{ad hoc} modeling assumptions.  These methods can be naturally extended to include high-dimensional inputs and outputs.  Such NN-based parameterizations of an experiment can serve many purposes.  One important example is the development of fast simulation tools like \textsc{DELPHES 3}~\cite{deFavereau:2013fsa}, where the detector response of all objects are parameterized.  NN-based parameterizations may provide more accurate models and require less person-hours to derive.  This is especially important for feasibility studies of future experiments where detailed studies are often not possible.  


The various loss functions may also be useful more generally, in order for the neural network training to be more tailored to the particular task and underlying data distribution.  In particular, proposals to use generative NN's for calorimeter simulations~\cite{Paganini:2017hrr,Paganini:2017dwg, ATL-SOFT-PUB-2018-001,Carminati:2018khv,Chekalina:2018hxi,Erdmann:2018jxd,deOliveira:2017rwa,Erdmann:2018kuh} are often trained with loss functions to include an absolute error term $\ell(E, E') = \abs{E - E'}$ for the true and reconstructed (generated) energies~\cite{Paganini:2017hrr,Paganini:2017dwg, ATL-SOFT-PUB-2018-001}. While this does encourage the reconstructed energy generally towards the center (median) of the response distribution, using the LGK loss and regressing towards the peak of the distribution may be more appropriate.


Another important example where response parameterizations are necessary is detector calibrations.  Supposing that $x$ is the true value and $y$ is the measured value of some observable, the most basic calibration approach would be to calibrate $y$ to $\nu(y)$, where $\nu$ is a NN trained to learn the central tendency of $x$ given $y$.  One problem with this approach is that the calibrated value depends on the distribution of $x$ used to train the NN~\cite{Cukierman:2016dkb}.  An alternative method that is used by ATLAS and CMS for jet calibrations is known as \textit{numerical inversion}, where the response parameterization of $y$ given $x$ is inverted: $y\mapsto \nu^{-1}(y)$.  Traditionally, this was done with fits and interpolation functions, but it can be generalized to neural networks where it can naturally depend on many features (\textit{generalized numerical inversion})~\cite{ATL-PHYS-PUB-2018-013}.  As the non-NN approaches typically use the mode, generalized numerical inversion with the LGK loss may be more appropriate than the MSE loss studied thus far.  One may further extend generalized numerical inversion to also include confidence intervals by using the quantile learning loss GAE in addition to only the measures of central tendency. 

\section{\label{sec:concl}Conclusions}

This paper has reviewed various strategies for machine learning different summary statistics of a detector response distribution $p(y|x)$ with neural networks including the mean, median, and mode for central tendency and the standard deviation and quantiles for measures of spread.  The summary statistic is linked to the loss function used in the neural network training.  While most of the loss functions have been combined with neural networks before, the authors believe that mode learning neural networks are presented here for the first time.  This is a particularly important case for detector parameterizations where the most likely value is often used instead of the mean or median to characterize the central tendency of the response distribution.  All of the methods presented here will expand the toolkit of available methods for detector characterization with a wide variety of applications in experimental particle physics and beyond.


\section*{Note added}

During the final preparation of this paper, a study from the CMS collaboration on $b$-jet calibration became public~\cite{CMS-PAS-HIG-18-027}.  This note used the Huber loss to characterize the central tendency and quantile learning to parameterize the resolution.  As far as the authors are aware, this is the first use of the latter technique in high energy physics.

\acknowledgments

This work was supported by the U.S.~Department of Energy, Office of Science under contracts DE-AC02-05CH11231 and DE-AC02-76SF00515.


\bibliography{mybib}

%
%
%
%
%
%
%
%
%
\end{document}